\documentclass[%
 preprint,
 amsmath,amssymb,
prb,
]{revtex4-2}

\usepackage{graphicx}% Include figure files
\usepackage{dcolumn}% Align table columns on decimal point
\usepackage{bm}% bold math
\usepackage{hyperref}% add hypertext capabilities
\usepackage{amsmath,amssymb,amsfonts}%
\usepackage{amsthm}%
\usepackage{mathrsfs}%
\usepackage{xcolor}%
\usepackage{textcomp}%
\usepackage{float}
\usepackage{ulem}
\usepackage{chngcntr}
\usepackage{appendix}
\usepackage{lipsum}% 
\usepackage{graphicx}%
\begin{document}

\preprint{APS/123-QED}

\title{3D Anderson localization of light in   disordered systems  of  dielectric particles}% Force line breaks with \\
%\thanks{3D Anderson localization of light}%

\author{Yevgen Grynko}
 \affiliation{BASF Coatings GmbH, Glasuritstraße 1, 48165 Münster, Germany}%Lines break automatically or can be forced with \\
 \email{yevgen.grynko@gmail.com}

\author{Dustin  Siebert}%
\affiliation{Department of Theoretical Electrical Engineering, Paderborn University, Warburger Str. 100, Paderborn, 33098, Germany}%

\author{Jan Sperling}
\affiliation{Theoretical Quantum Science, Institute for Photonic Quantum Systems, Paderborn University, Warburger Str. 100, Paderborn, 33098, Germany}%

\author{Jens Förstner}
\affiliation{Departmentt  of Theoretical Electrical Engineering, Paderborn University, Warburger Str. 100, Paderborn, 33098, Germany}%

\date{\today}% It is always \today, today,

\begin{abstract}

We investigate light transport in three-dimensional disordered media composed of irregular dielectric particles using large-scale full-wave simulations. For subwavelength particles with size parameter $kr \approx 1$ and high refractive index contrast, we observe a transition from diffusion to a regime characterized by non-exponential decay of time-resolved transmission as disorder increases. The corresponding time-dependent diffusion coefficient decreases with time and approaches a $t^{-1}$ scaling at long times. This dynamical slowdown is accompanied by the emergence of spectrally isolated transmission resonances with Thouless conductance below unity, indicating the dominance of long-lived modes with weak spectral overlap. The late-time near-field maps reveal evolving, non-propagating clusters of intensity hotspots.  Together, the transport, spectral, and near-field signatures provide consistent numerical evidence for Anderson localization of light in three-dimensional disordered dielectric media.
  
\end{abstract}

\maketitle

%\tableofcontents

\section{Introduction}\label{sec1}
A complete understanding and control of light propagation in discrete disordered media can be advantageous in different applications in photonics, like imaging and focusing in random media, random lasing, radiative transfer and optical remote sensing \cite{Jeong2018, Wiersma2008, NELSON2018483, SHKURATOV2002396, Jacques_2013}.  However, modeling light transport phenomena in highly disordered turbid media  remains a challenge due to the inherent complexity and multi-scale nature of the problem. While classic wave diffusion and weak localization, observed as  coherent backscattering enhancement \cite{Akkermans1988, AEGERTER20091, Kaveh1986}, are well understood, strong Anderson localization of light (AL) in 3D structures is a subject of ongoing research.
AL has been predicted for media with strong disorder and strong multiple scattering \cite{Anderson1985,Tiggelen2000}. Generally, it is characterized by a transition from diffusive propagation to a reduction and even complete halt of diffusion. An explanation involving trajectories with closed loops and increased return probability is often suggested \cite{Cherroret2008,Naraghi_etal2016, Wiersma1997}.  However, the actual mechanism behind such dynamics has not been described yet in detail.  The phenomenon is general in nature and can be observed for different kinds of waves, e.g., in acoustics \cite{Hefei2008}, quantum systems \cite{Kondov2011, Crespi2013}, 2D  \cite{Riboli2011, Schwartz2007} and spatially correlated 3D photonic structures \cite{Haberko2020}.  The existence of  AL for electromagnetic waves in 3D  uncorrelated media and metallic materials has been demonstrated  recently through a numerical study \cite{Yamilov_etal2022}.
%for metal overlapping spheres whereas analogous dielectric structures did not exhibit the phenomenon. 
  
Observable characteristics  of AL are the reduction of transverse spreading of the propagating light beam \cite{Yilmaz2019, WANG2021,Yamilov_etal2022} and, in particular, time-delayed transmitted energy decay \cite{SkipetrovVanTiggelen2006,Yamilov_etal2022}. Transmission spectrum in this case is characterized by non-overlapping sharp peaks indicating long-lived mode domination at longer times \cite{Chabanov2000, THOULESS1977, vanRossum1999, Mondal2019}. Numerical modeling of light transmission through layers of overlapping metallic spheres revealed these  anomalies as unambiguous signatures of  AL \cite{Yamilov_etal2022},  %of light transmission through 3D dense layers  of perfect electric conductor (PEC) overlapping spheres 
 suggesting its general possibility in 3D. At the same time analogous dielectric structures did not exhibit the phenomenon \cite{Yamilov_etal2022}. Thus, the question whether AL can be observed  for electromagnetic waves and disordered dielectric 3D media remains an open problem to date. Experimental measurements of light transmission by layers of highly scattering particulate materials could shed light on the problem \cite{Wiersma1997,Stoerzer2006,Sperling2013}. However, certain problems with precise control of the sample properties and accounting for the effects of absorption \cite{Scheffold1999} and inelastic scattering \cite{Scheffold2013,Skipetrov_2016,Sperling_2016} make this difficult. 

An important condition for the AL emergence is the dominance of coherent multiple scattering that can be expressed through the heuristic Ioffe-Regel criterion \cite{IoffeRegel1960}. It was originally formulated in terms of the scattering mean free path $l_s$, expressing the breakdown of wave propagation when the wavelength becomes comparable to the distance between successive scattering events, i.e.  $kl_s \lesssim 1$ \cite{Anderson1985, SkipetrovVanTiggelen2006, SkipetrovSokolov2018, Abrahams1979, Wiersma1997, Cherroret2008, vanRossum1999}. Within the framework of transport theory, however, an analogous condition involving the transport mean free path $l^*$ was later introduced \cite{SHENG1995, LAGENDIJK1996, Mondal2019, Haberko2020, Sperling2013, Sperling_2016}, where $l^*$ accounts for angular redistribution of energy and is therefore more directly connected to diffusive transport. The two quantities are related through the single-scattering anisotropy factor  $g$ as $l^* = l_s / (1-g)$ \cite{ISHIMARU1978}.  As a result, both formulations coexist in the literature, and the term “Ioffe–Regel condition” is  used with either $l_s$ or $l^*$. In many cases, authors adopt one of these parameters and the rationale for this choice is not discussed explicitly, despite the fact that $l_s$ and $l^*$ can differ substantially. This ambiguity complicates quantitative comparisons between studies and highlights the need to clearly specify which definition is employed when discussing the approach to localization. 

In this work, we apply numerical simulations and high-performance computing to demonstrate that  AL signatures can be observed in light transmission by  3D  layers of irregular dielectric particles. Reaching high enough degree of spatial disorder, we obtain a transition of classical diffusion characterized by the exponential time dependence of transmission $T(t)$ to  $T(t)$ deviating from an exponent and time-dependent decay rate $D(t)$ being in agreement with self-consistent theory of localization  \cite{SkipetrovVanTiggelen2006} . This behavior is accompanied by the emergence of  isolated sharp peaks in the transmission spectrum.  %We adopt the transport-based formulation and evaluate $kl^*$ independently from time-resolved observables, providing a consistency check for the approach to the localization regime.

We note that the problem requires proper description of the disordered geometry of a model medium and careful accounting for the multiple scattering near-field effects, including the polarization state of the scattered near field  \cite{Grynko2020, Naraghi_etal2015, Naraghi_etal2016, Astratov2007, Skipetrov2005}.  In existing analytical solutions, simplified descriptions of the media are used (e.g., \cite{Skipetrov2005, Cherroret2008, TiggelenSkipetrov2021})  which   can be insufficient if applied to the vector electromagnetic field in real 'white paint'-like 3D media. With full wave numerical simulations one can avoid approximations and consider pure electromagnetic scattering in a target structure with arbitrary geometry   \cite{Grynko2020,Grynko_chapter_2022,GRYNKO2022Icarus, Yamilov_etal2022, Conti2008, WANG2021, Haberko2020}. 

%%%%%%%%%%%%%%%%%%%%%%%%%%%%%%%%%%%%%%%%%%%%%%%%%%%%%%%%%%%%
\section{Numerical model}\label{sec2}

We use the discontinuous Galerkin time-domain method (DGTD) \cite{HESTHAVEN2002, Busch_review_2011} to solve Maxwell's equations for finite and periodic multi-particle layers illuminated by a plane wave  as well as by a focused beam. 
 All simulations were performed using a self-developed light scattering code based on the open source DGTD Maxwell solver MIDG  \cite{midg_code}. The \textit{CST Microwave Studio} software package based on the Time Domain Finite Integration Technique was used for a qualitative benchmarking the DGTD simulation results for simplified setups (Suppementary material \ref{CST_check}). In our code, we implemented plane-wave and focused beam lights sources, boundary conditions and material description \cite{Grynko_chapter_2022, Grynkochapter2017}. A plane wave source is realized using the total field/scattered field technique. For simulations with a focused beam we implemented the plane wave superposition method \cite{Capoglu2008}. An open computational domain is represented using the PML absorbing  boundary condition. For layers infinite in transverse directions, periodic boundary conditions (PBC) were used. For the time-resolved transmission measurements we illuminated  target layers with a linearly polarized short plane wave pulse. The intensity of the transmitted signal was recorded, correspondingly, in a monitor plane behind the layer. The intensity curves in all plots are normalized by the peak intensity of the transmitted pulse. 
 
 The Ioffe-Regel criterion unconditionally requires dense packing of scatterers. Maximally random jammed packings  of regular constituents, like spheres and regular polyhedra, always form hyper-uniform structures regardless of the packing algorithm \cite{Torquato2003}. This kind of discrete media packed near their corresponding maxima  have a reduced degree of disorder which  influences their transmission properties (e.g., \cite{Cheron_etal2022, Grynko2020} ). The problem can be solved by considering scatterers with random irregular shapes.  Therefore, we use sets of particles  created with a self-developed shape generator based on the Gaussian random field (GRF) approach \cite{Grynko2018, Grynko2020, Grynko_chapter_2022}.  For generation of the particulate samples with different dimensions and controlled filling fractions we use the \textit{Bullet} Physics engine \cite{coumans2021}. This is an open-source C++ library that can be also used as an add-on in the Blender software. It allows simulation of dynamics and collision of multiple arbitrary 3D shapes in time domain. Thus, simulating natural powder mechanics samples with realistic topologies can be created. For sparse distributions of the pre-generated random irregular shapes we simulated their free fall  on a substrate in a closed  volume.  3D models of the generated layers were then used for tetrahedral mesh generation. The highest refractive index of the non-absorbing dielectric material was $n=3.0$ and the spatial resolution in the corresponding regions was $\sim15$ tetrahedral cells per central wavelength of the incident pulse. The nodal expansion order in the DGTD numerical scheme was $N=3$.  All  the DGTD simulations were done on the HPC cluster Noctua 2 of Paderborn University using up to 30 cluster nodes per simulation. 

We simulate light transmission by mono-disperse  layers with up to 20000 GRF particles. The constituent particle size is determined by the goal of reaching  the strong scattering regime. In particular, we consider particles with characteristic size $X_r = kr \approx 1 $, where $r$ is the radius of the circumscribing sphere, and operate at volume fractions approaching random close packing, up to $\rho = 0.5$. Under such conditions, both the structural length scale (particle size and inter-particle spacing) and the wavelength become comparable, which generically leads to a scattering mean free path on the order of the wavelength,  $k l_s \sim 1$. This choice is further supported by recent numerical evidence of localization in 3D metallic slabs with spherical particle size  $kr \approx 1$ \cite{Yamilov_etal2022}. While the exact correspondence between metallic and dielectric systems is not direct, these results identify this size as a starting point for exploring the breakdown of diffusive transport. The corresponding transport parameter $kl^*$  is subsequently determined from time-resolved transmission measurements (section \ref{PBC_results}), providing an internal consistency check of the achieved transport regime. We note that the Ioffe–Regel condition is employed here as an order-of-magnitude indicator of strong scattering rather than a strict localization criterion  \cite{SkipetrovSokolov2018}.

\begin{figure}[h]%
    \centering
    \includegraphics[width=10cm]{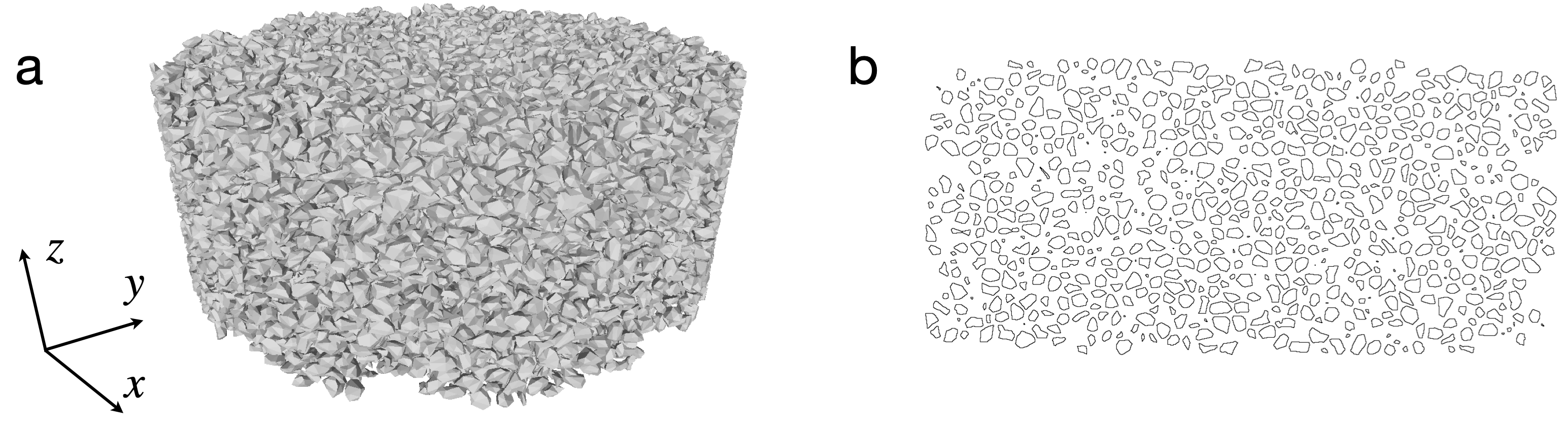}
    \caption{ A layer of 19000 irregular particles packed with a volume fraction of 0.44 (a) and its vertical cross-section (b).}
    \label{perturbed_sample}
    \end{figure}

 We consider two types of finite cylindric layers and layers with periodic boundaries in transverse directions. Cylindric symmetry of finite samples may, in principle, affect the result. Therefore, we use samples with two types of the  side boundaries: a randomly perturbed one and a smooth boundary. In the first case,   defects of the scale of the wavelength $\sim \lambda$,  breaking such symmetry, are introduced (Fig. \ref{perturbed_sample}).  In the second, case a smooth boundary forms a dielectric discontinuity  between the disordered medium and free space and, generally, one can expect whispering-gallery (WG) modes due to total internal reflection and cylindric symmetry \cite{McCall1992}.  
 
\section{Results and discussion}\label{sec3}

%%%%%%%%%%%%%%%%%%%%%%%%%%%%%%%%%%%%%%%%%%%%%%%%%%%%%%%%%%%%%
\subsection{Transmission by finite cylindric samples}

\begin{figure}
\centering
\includegraphics[width=11cm]{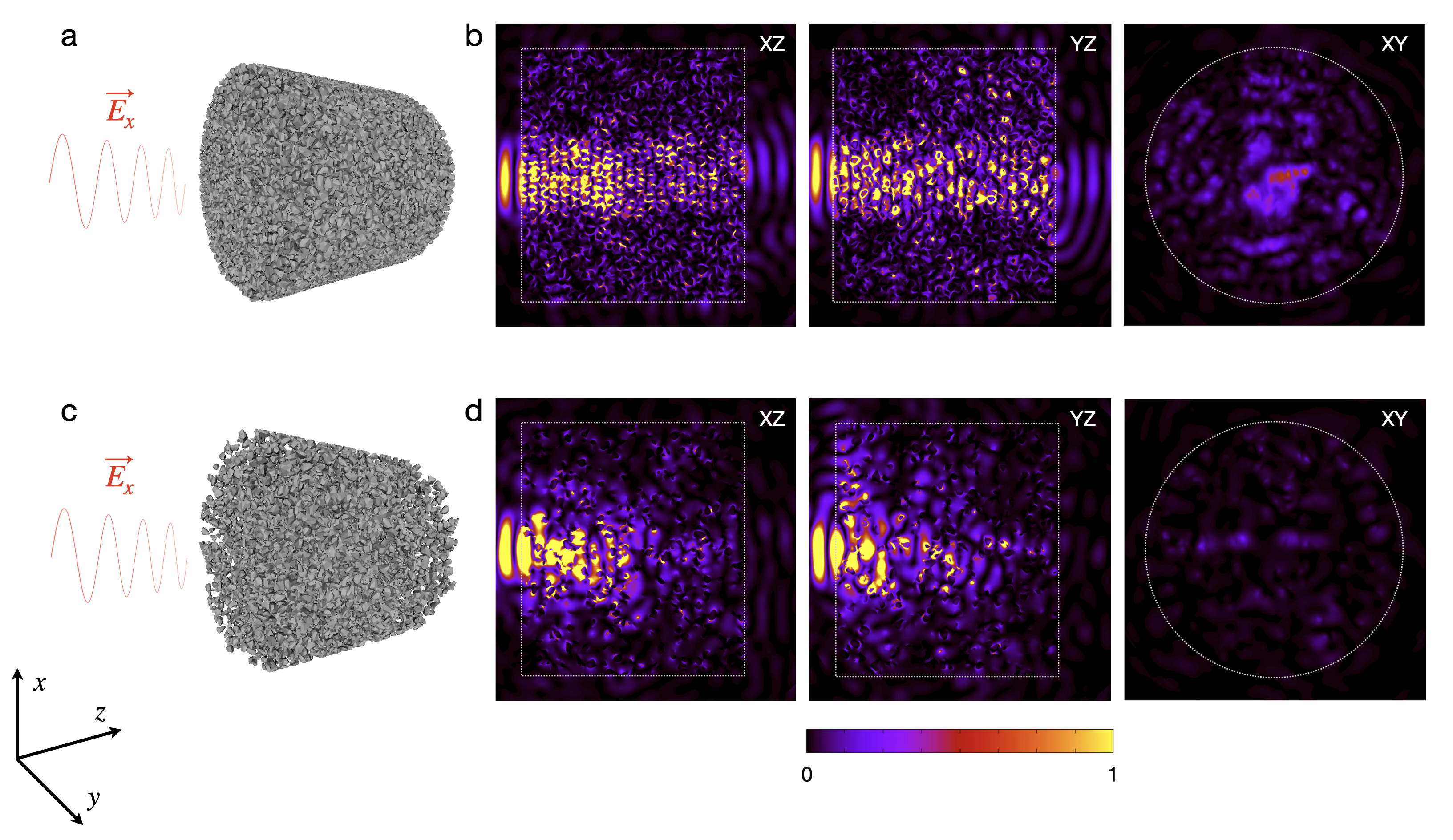}
\caption{Propagation of a continuous, focused beam through a dense and a sparse layers of irregular particles. (a) and (c), samples with volume fractions of 0.48 (dense) and 0.16 (sparse), respectively. (b) and (d), steady-state near-field intensity distributions $|E|^2$ in XY, YZ and back-XZ planes for both layers.
The incident beam is $E_x$-polarized. The size of particles is $X_r=1.1$, and the refractive index is $n=3.0$. 
}
\label{fig1}
\end{figure}

\begin{figure}[h]%
\centering
\includegraphics[width=11cm]{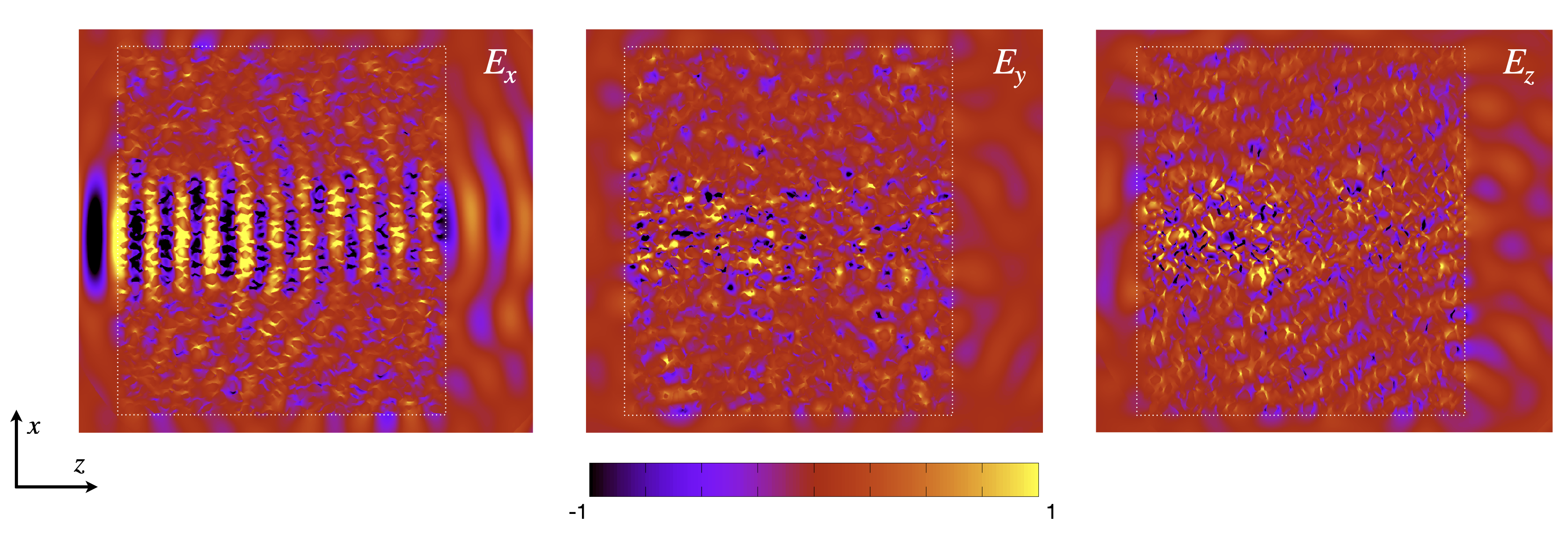}
\caption{Steady-state distributions of the three components of the electric field at the propagation of a continuous focused beam through the dense layer  with $\rho=0.48$ and $n=3.0$ shown in Figure \ref{fig1}a.}
  \label{fig2}
\end{figure}

In this section present the results for finite samples probed by focused-beam with analysis of the propagation and plane-wave  pulses excitations for time-resolved transmission measurements. 

  AL is expected for materials with high refractive indices. E.g., TiO$_2$ powders are reported as those where transverse localization and anomalous transmission can be observed \cite{Sperling2013, Stoerzer2006}. Therefore, we first fix the refractive index at $n=3.0$ to study the role of the volume fraction factor. 

Classic diffusive propagation of light in sparse media with widely separated particles ($l_s > \lambda $)  is characterized by a random walk driven by multiple scattering and diffraction from individual scatterers. These interactions create a diffuse energy distribution where light spreads in both forward and transverse directions, with the degree of forward-peaking defined by the anisotropy factor $g$ \cite{ISHIMARU1978}. Consequently, the light direction becomes increasingly decoupled from the initial source with each subsequent scattering event. The geometry of the field spread experiences a change when particles are densely packed. Generally, at dense packing, far-field scattering becomes impossible, and classic multiple scattering turns into percolation of light along random transmission channels \cite{Grynko2020, Naraghi_etal2015, Astratov2007}. This can lead to formation of Caylee-tree-like channels (e.g., \cite{Grynko2020}) or a transversely localized propagation \cite{Yilmaz2019, WANG2021}, depending on the  particle size and volume fraction.

These two cases are demonstrated in  Fig. \ref{fig1}, where we study focused beam propagation through thick layers of  particles with  size $X_r = 1.1$ and  refractive index $n=3.0$. The dimensionless size parameters of the diameter and thickness are $X_R = 22.1$ and $X_L = 19.5$,  respectively. The internal field patterns are compared for a dense and a sparse structures of $\rho = 0.48$ (Fig. \ref{fig1}a) and 0.16 (Fig. \ref{fig1}c),  respectively.  The number of particles in the dense sample is 20000. A continuous $E_x$-polarized focused beam incident on the front side in each case propagates in $Z$ direction. Steady-state intensity distributions as cross-sections in different planes are shown in Figs. \ref{fig1}b and \ref{fig1}d. The "XY" panels correspond to the cross-sections near the back sides of the samples.

In the sparse case (Fig. \ref{fig1}d), we observe conventional multiple scattering by particle groups and individual particles. In a dense system, the size of voids is reduced to the scale of the wavelength or smaller, and a focused beam tends to preserve its diameter along the path, re-emitting transmitted energy in the center of the back side.  Isotropic spreading occurs due to the leakage of energy through the evanescent field coupling. Therefore, the intensity of the beam decreases, but its geometric cross-section is preserved at the distance of the layer thickness.  This behavior reflects suppressed transverse spreading rather than ballistic transport and occurs simultaneously with the formation of a spatially extended, randomized field component.

Additional  characterization  can be obtained from the analysis of the polarization state of the near field. A steady-state distribution of the electric field component $E_x$  in Fig. \ref{fig2} computed for the same dense layer and same conditions as in Figure \ref{fig1}b shows a partial conservation of  the  original polarization state of the propagating beam at the exit. However, the $E_y$ and $E_z$  components are excited along the path and play a role in the field coupling process. All three components  of the randomized field that uniformly fills the rest of the sample volume have comparable amplitudes which indicates high degree of depolarization. 

The beam propagation dynamics can be better seen in Video \ref{short_pulse_video} of the Supplementary material \ref{Supplement_video}. It shows percolation of a  short pulse  through a dense layer. Light propagates  mainly through evanescent coupling between individual scatterers without conventional free-space transport of the scattered waves. This behavior can be qualitatively viewed as analogous to the evolution of a cellular automation (CA) in which the vector field state in an elementary cell at each time step is determined by its neighborhood, i.e. the shape of the neighbour particles or the shape of a sub-wavelength void that is formed by the neighbors. The laws of interaction of the vector field and the air-dielectric interface together with interference become the CA rules. 
%Then, a sequence of such interactions determines the global evolution of the  vector field from the initial state. 
This analogy emphasizes the importance of short-range interactions in shaping the global field dynamics at dense packing and high refractive index. The complex dynamics of the randomized diffusive component at early time after the short pulse propagation in shown in Video \ref{short_plane_wave_video}.

The absence of pronounced transverse spreading in a dense sample does not imply a ballistic regime. Instead, it reflects the dominant role of near-field interactions at high packing densities and refractive index contrast. Energy is continuously redistributed among neighboring scatterers, leading to the excitation of the $E_y$ and $E_z$  field components and the formation of a volume-filling scattered field that dominates the long-time dynamics and gives rise to the exponential tail in $T(t)$. Thus, the system remains non-ballistic and strongly scattering, even though the spatial extent of the transmitted intensity profile is preserved. This coexistence of contributions with different decay characteristics is consistent with the general picture of wave transport in strongly scattering media \cite{Akkermans_Montambaux_2007, Yoo90}.

From the above analysis, we see that the fundamental property of the observed transport mechanism in a densely packed layer is concentration of the transported energy  in the free-space cavities, while less is refracted inside the dielectric particles. A wave can couple to the neighbor-particle interface while being still coupled to the previous one propagating in a void between two interfaces. On the other hand, these interfaces are  disconnected and belong to different scatterers. Thus, the field appears to be constantly coupled and the concept of discrete scattering events, as well as the term "free path length", becomes ill-defined. 
%Therefore, the term "free path length" becomes irrelevant in its classic definition. 
The condition  $kl_s \lesssim 1$ implies also an extremely small mean free  time $t_s$, and the near field, being coupled to a particle, most often reaches its neighbour at time $t \ll T$, where $T$ is the wave oscillation period. Thus,  the Ioffe-Regel criterion for $l_s$ should be considered as qualitative for the localization problems in consistence with theoretical results \cite{SkipetrovSokolov2018}. The parameter $l_s$ can be measured only indirectly and the existing methods for such indirect estimations should be, therefore, validated for the case of the vector field propagating in a very dense 3D medium. This is, however, beyond the scope of this paper. 

  \begin{figure}[h]%
    \centering
    \includegraphics[width=10cm]{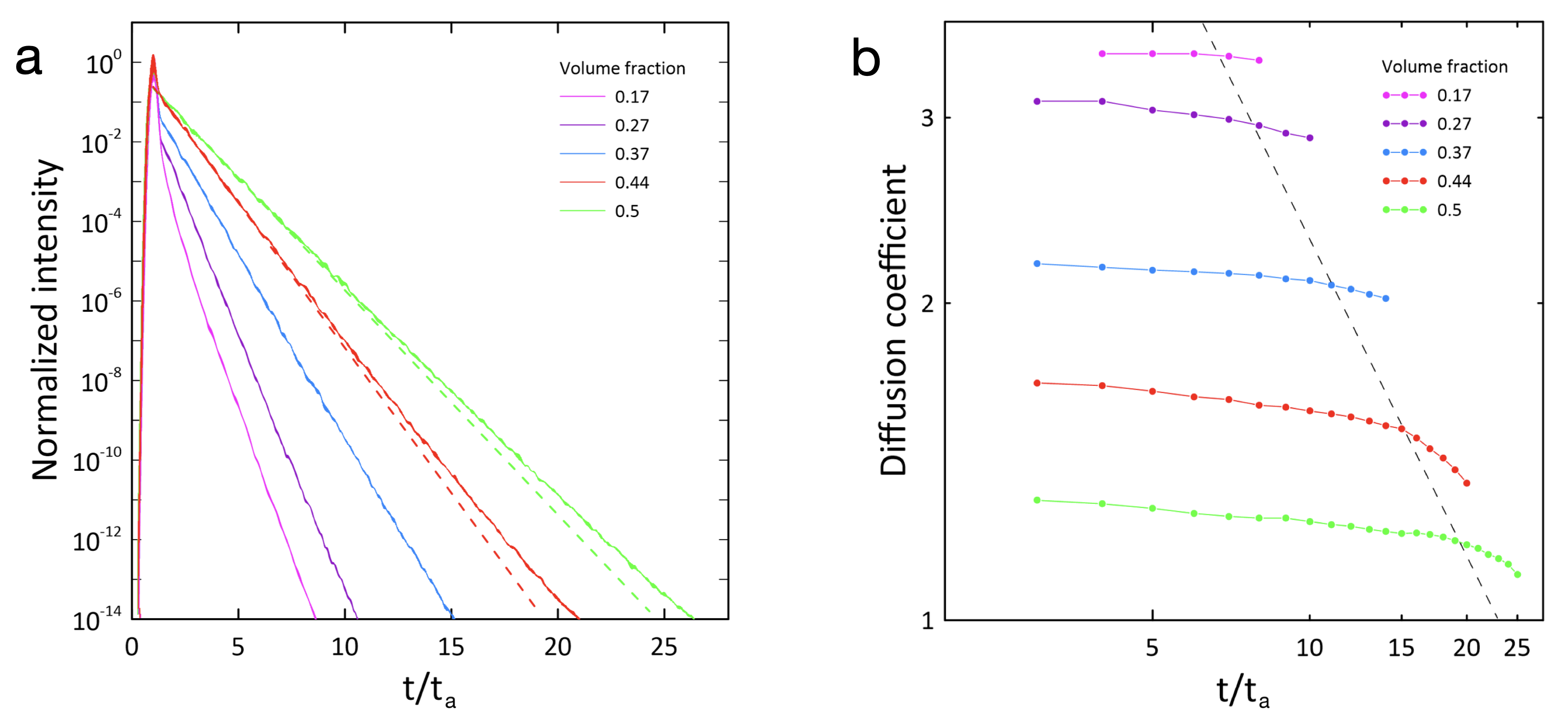}
    \caption{ (a) Normalized transmission of a short pulse by layers with perturbed side boundaries and volume fractions from $\rho=0.17$ to $0.5$ as a function of time $T(t)$. The size of particles is $X_r=1.1$ and the refractive index is $n=3.0$.  (b) Diffusion coefficient $D(t)$ obtained by local exponential fitting of $T(t)$. Both axes have logarithmic scale. Dashed line shows a  $t^{-1}$ dependence. 
    }
    \label{fig3}
    \end{figure}
 
A conventional scaling analysis based on total transmission as a function of sample thickness is, in principle, a standard approach to identify localization \cite{Wiersma1997}. In our system, however, a directionally transmitted component can persist across the slab and thickness dependence does not isolate intrinsic transport properties unless much larger samples are used to suppress this contribution. Instead, we analyze the time-resolved transmission, whose long-time behavior is dominated by the randomized component and directly reflects the transport regime relevant for localization.  For this, we generated layers with various volume fractions $\rho$ and larger diameter to thickness ratios, i.e., horizontally more extended, and numbers of particles up to 19000. 

 At first, we consider samples with perturbed side boundaries and with dimensions of $X_R = 32$ and $X_L = 13$. The volume fraction $\rho$  is varied  between 0.17 and 0.5 and the refractive index is fixed at $n=3.0$. An example of a dense sample with $\rho=0.44$ and its cross-section is shown in Figures \ref{perturbed_sample}. The computed transmissions $T(t)$ for different volume fractions  are shown in Fig \ref{fig3}a. The layers are illuminated by a short Gaussian pulse. The arrival of the transmitted pulse intensity peak at the detector plane is recorded as time $t_a$ for each case. The horizontal axis represents  normalized time $t \slash t_a$. The maximum observation time is limited to $ \approx 25 t_a$ as the samples are relatively thin and the measured intensity quickly reaches the numerical accuracy of double-precision numbers. However, we observe a non-exponential dependence of $T(t)$ for larger volume fractions   near this limit.  With increasing turbidity at larger $\rho$, we approach the subwavelength scale of inter-particle spacing obtaining not only a decreasing slope of $T(t)$ but a noticeable deviation from exponential behavior longer times. 
   
 Non-exponential decay rate can be formally represented by means of a time-dependent diffusion coefficient $D(t)$. At the localized state, it decreases as  $t^{-1}$ according to the solution of the diffusion equation for the case of $kl_s \lesssim 1$ in time domain  \cite{SkipetrovVanTiggelen2006, Yamilov_etal2022}. Figure \ref{fig3}b shows time-resolved diffusion coefficients $D(t)$ deduced from the computed $T(t)$ by local exponential fitting. One can see a transition from small variation of $D(t)$ in sparse structures to a steep  function at $\rho > 0.44$ and times $t > 15t_a$. 
   
\begin{figure}[h]%
  \centering
  \includegraphics[width=11cm]{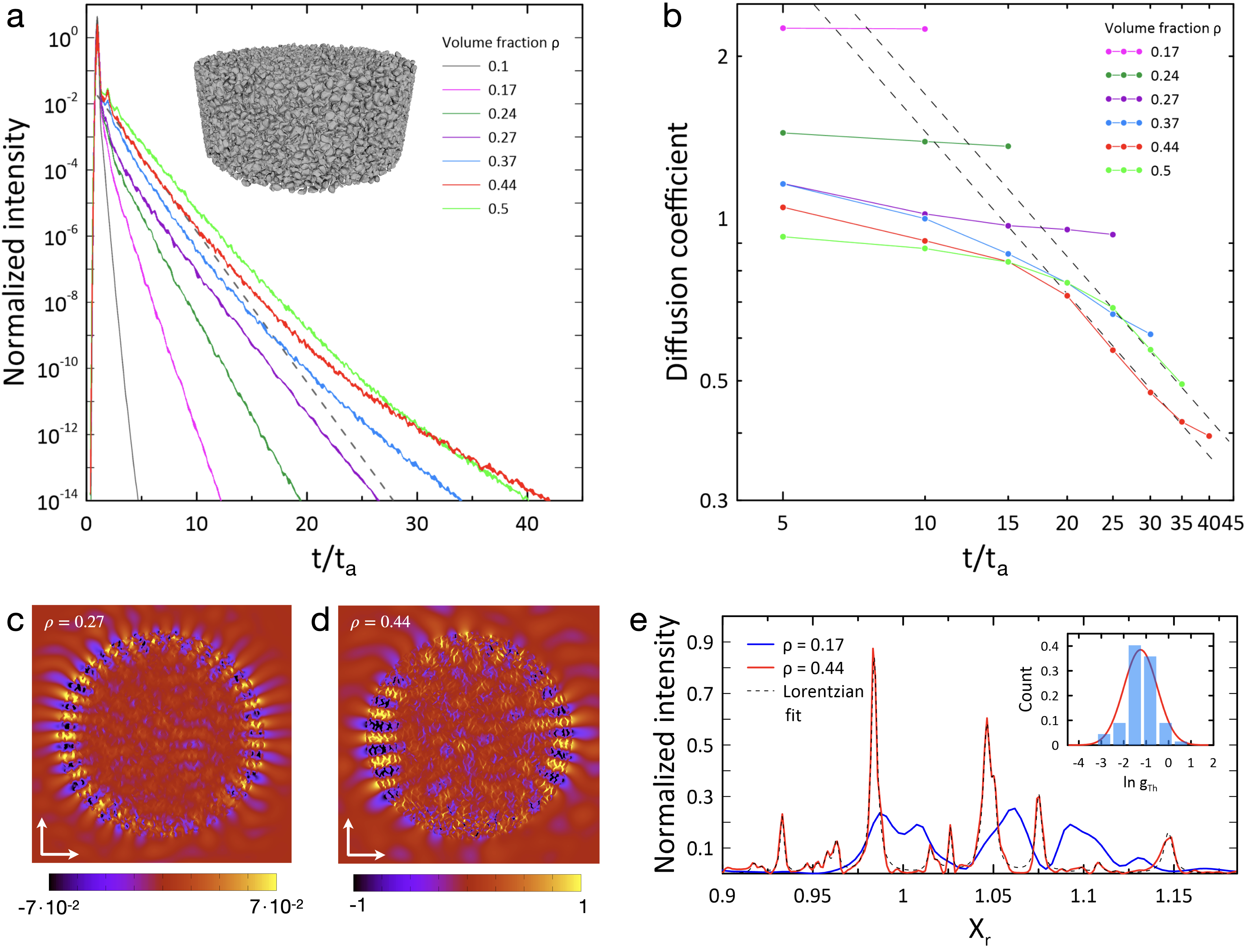}
  \caption{Simulation results for layers with different volume fractions and smooth side boundaries.  The refractive index of the material is $n=3.0$. (a) Normalized transmission of a short pulse by layers with volume fractions  from $\rho=0.1$ to 0.5 as a function of time $T(t)$. 
   (b) Time-dependent diffusion coefficient $D(t)$ obtained by local exponential fitting of $T(t)$. Dense layers demonstrate a $t^{-1}$ (dashed lines) dependence at times $t > 20t_a$.  (c) and (d), Snapshots of the  internal electric field (XY-cross-sections, $E_x$ component) distributions showing a whispering-gallery wave propagating along the boundary at time $t = 25 t_a$ for a moderately sparse ($\rho=0.27$) and a dense ($\rho=0.44$)  layers. 
   %In the last case $T(t)$ is exponential despite the whispering-gallery effect. 
   (e) Normalized spectra of transmitted light as functions of the dimensionless size parameter $X_r$ (inverse wavelength expressed relative to the particles size) for sparse $\rho=0.17$ and dense $\rho=0.44$ layers. Dashed line corresponds to a Lorentzian fit for $\rho = 0.44$. The inset shows the Thouless conductance distribution $P(\textrm{ln} ~ g_{Th})$ for  $\rho=0.44$  that has a mean value $\mu = -1.25$ and a standard deviation $\sigma = 0.75$. % and approximately follows a Gaussian law.
   }.
  \label{fig4}
  \end{figure}

 Evidently we would have to extend the observation time limit in order to study the evolution of  $T(t)$ at longer times and to find out if $D(t)$ can reach the predicted $t^{-1}$ dependence. This could be achieved by increasing  sample dimensions roughly by a factor of two or more and, correspondingly the number of particles to $N \gtrsim 10^5$, which makes the problem computationally extremely expensive, even for modern large scale HPC clusters. Another way to hinder energy decay in time is to create conditions for a whispering-gallery effect at the side boundary of a cylindric layer. For this purpose we did computations for layers with smooth side boundaries. The results for  $T(t)$ and $D(t)$ dependencies at different volume fractions from $0.1$ to $0.5$ are shown in  Figures \ref{fig4}ab. The dimensions of the layers are $X_R = 28$ and $X_L = 12$. A densely packed sample with $\rho = 0.44$  is shown in the inset in Figure  \ref{fig4}a.  Figures \ref{fig4}cd represent snapshots of the electric field distributions ($E_x$  component) in the horizontal XY-plane  at time $t = 25t_a$ for a dense ($\rho = 0.44$) and a moderately sparse ($\rho = 0.27$) layers.  The WG waves, excited by the incident pulse, allow us reaching   $t \approx 40t_a$ for the samples with large volume fractions and relatively small thickness.  At the same time the field distributions remain random in the bulk. Despite the WG effect, present in all cases, including  the sparse structures, they show  different transmission properties. The $T(t)$ plot in Figure \ref{fig4}a clearly demonstrates an emergence of the non-exponential tail from an exponential decay at large  volume fractions $\rho \gtrsim 0.3$ and times $t > 20t_a$.  %The WG effect  is observed in all the layers above and below  this value,  
   
 The diffusion coefficient  $D(t)$ at times   $t  \gtrsim 20t_a$  obtained for the dense structures with $\rho \ge 0.44$  follows a $t^{-1}$ fit in Figure \ref{fig4}b indicating a superposition of contributions from multiple modes with different decay rates which is consistent with AL regime. This result is in agreement, in terms of time and the volume fraction threshold, with that obtained for layers of metallic spheres in \cite{Yamilov_etal2022}. Interestingly, a WG mode is  established at times $t \approx 10t_a$, prior to the  emergence of the $D(t) \sim t^{-1}$ regime.  
 This  proves that it is not the WG mode itself but the particle packing density which is causing the non-exponential behavior of $T(t)$.

  To make the characterization of the localized transport  more complete we compute transmission spectra for the sparse ($\rho = 0.17$) and dense ($\rho = 0.44$) samples  (Figure \ref{fig4}e). For the horizontal axis a dimensionless size parameter $X_r$ is used, which is equivalent to the inverse wavelength relative to the particle size. The $X_r$ range in the plot spans approximately the  full width at half maximum of the incident pulse. The $\rho = 0.17$ spectrum is represented by broad, overlapping, maxima. The spectrum for $\rho = 0.44$ consists of  non-overlapping sharp peaks resulting from the states spatially confined in the medium   \cite{Riboli2011, Chabanov2000, THOULESS1977, vanRossum1999, Mondal2019}. Dashed line shows a spectral fit obtained with an ensemble of Lorentzian oscillators.  The separations of peaks are larger than their widths, which indicates that the Thouless criterion for AL is satisfied. With this, a Thouless conductance $g_{Th}$
  can be defined as the ratio $g_{Th} = \delta \omega / \Delta \omega$, where $\delta \omega$ is the average width of two adjacent modes in the spectrum and $\Delta \omega$ is their spectral separation. Thus, $g_{Th} < 1$ corresponds to a localized propagation. In the case of $\rho = 0.44$, where the signatures of localization emerge, we estimate it as an average over all visible modes as $<g_{Th}> \approx 0.36$. The inset in Figure \ref{fig4}e shows the distribution  $P(\textrm{ln} ~ g_{Th})$. It approximately follows a Gauss law according to the theoretical predictions \cite{vanRossum1999, Mondal2019}.

  \begin{figure}[h]%
  \centering
  \includegraphics[width=10cm]{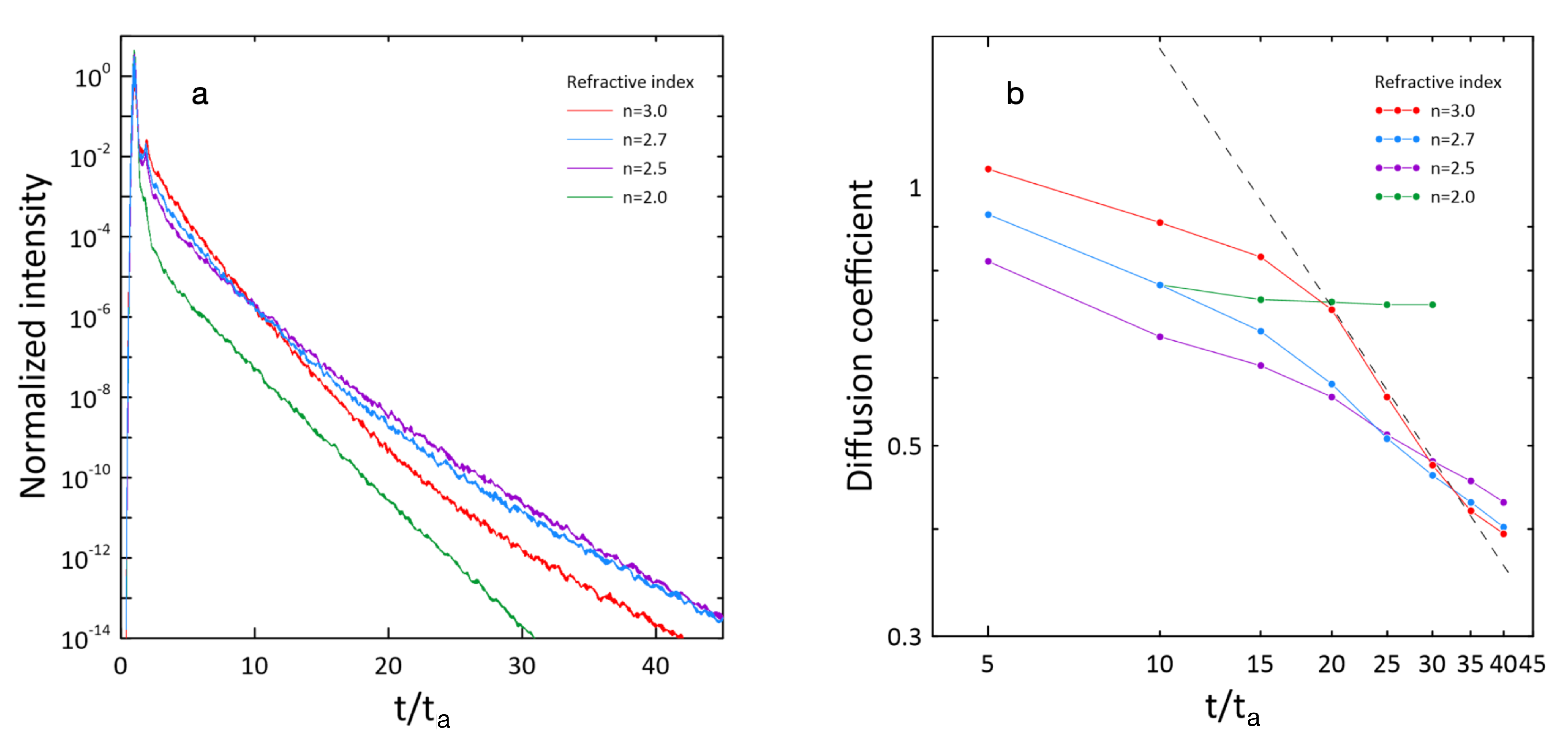}
  \caption{(a) Normalized transmission of a short pulse  through a layer with a volume fraction $ \rho=  0.44$ and a smooth boundary as a function of time $T(t)$ at different refractive indices. (b)  Diffusion coefficient $D(t)$ as obtained by a local exponential fitting of $T(t)$. Dashed line shows a  $t^{-1}$ dependence. 
  }
  \label{fig5}
  \end{figure}

  Another free parameter of the medium that is important for light transport is the refractive index of the material. To illustrate the transition between diffuse and localized regimes from this point of view, we do similar computations of $T(t)$ for the dense sample with $\rho = 0.44$ and refractive indices ranging from $n=$2.0 to 3.0. Figure \ref{fig5} shows that $n$ must be larger than 2.5 for a non-exponential decay, and further increase of $n$ enhances this effect at longer times. This is also consistent with theoretical estimations and experimental measurements for materials with different $n$ that suggest $n \gtrsim 2 $    \cite{Anderson1985, Sperling2013}. Thus, the refractive index contrast is another critical factor for AL in dielectric structures that can impede or enhance  the localized behavior. It turns out that $n$ should not be necessarily extremely large and a continued experimental search for AL in 'white paint' powders  like TiO$_2$ is not futile. 

\begin{figure}%
\centering
\includegraphics[width=11cm]{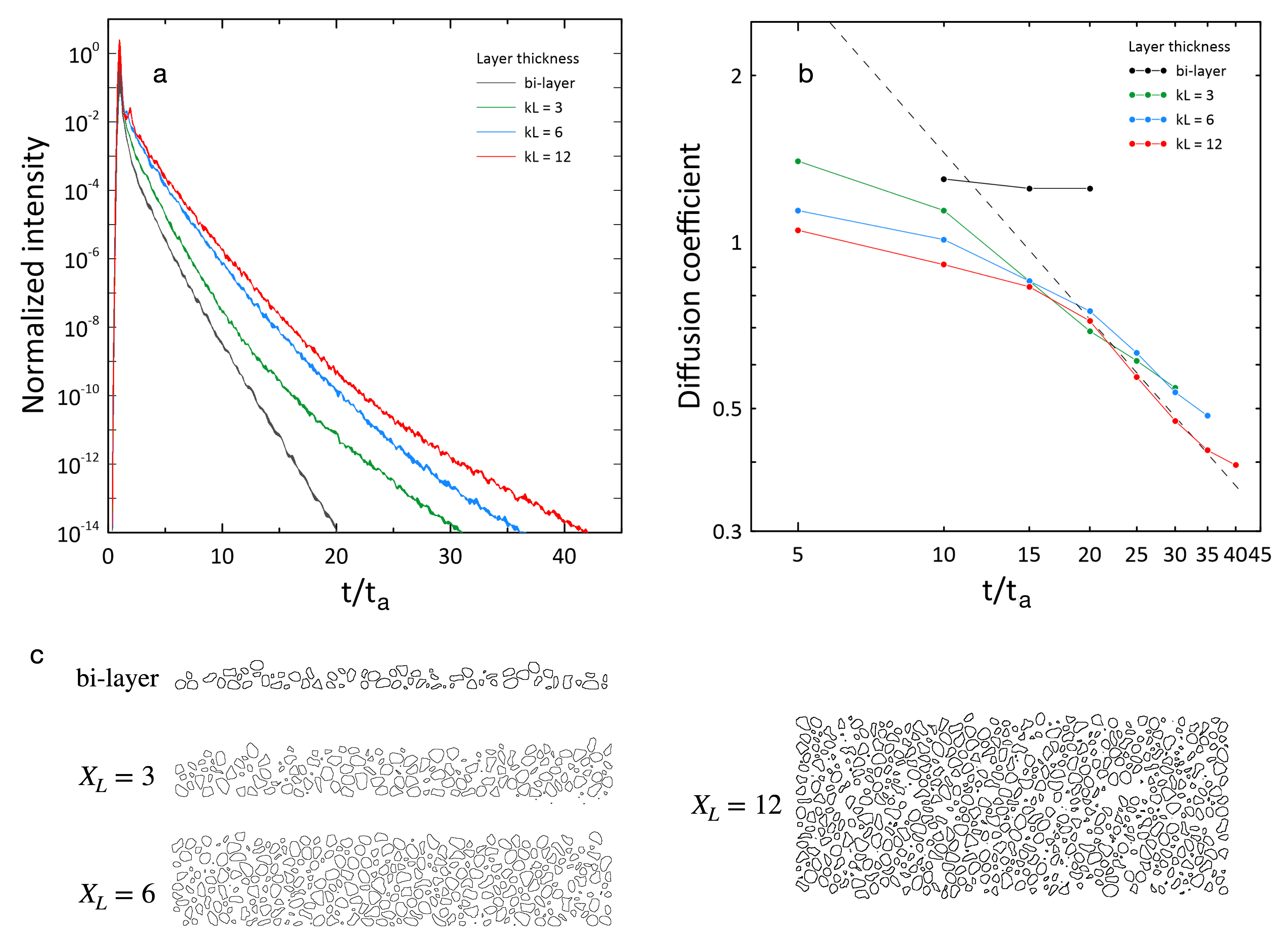}
\caption{(a) Normalized transmission of a short pulse  through layers with cylindric symmetry and different thickness $X_L$ as a function of time $T(t)$.  The diameter and volume fraction are  $X_R=26$ and  $ \rho= 0.44$, correspondingly,  and the refractive index is $n=3.0$.  The energy decay from a simple double layer of particles is exponential while thicker layers demonstrate increasingly non-exponential transmission. (b) Diffusion coefficient $D(t)$ obtained by local exponential fitting of $T(t)$. $D(t)$  evolves from a constant to a $t^{-1}$ fit (dashed line) with increasing thickness. (c) Vertical cross-sections of the samples with different thickness.}
\label{different_thickness}
\end{figure}

To study layer thickness dependence, we take the sample with $X_R=26$, $X_L = 12 $ and $\rho=0.44$ as a reference and consider reduced structures with the same bulk volume fraction and thicknesses $X_L = 6$ and 3 and an elementary bi-layer structure. The refractive index is $n=3.0$. The results of the simulated transmission $T(t)$ for four layers are presented in Figure \ref{different_thickness}a. In all four cases a WG mode is excited allowing us reaching times  $t > 20 t_a$.  A bi-layer shows  exponential decay despite the WG effect. The case of $X_L = 3$ is equivalent to a doubling the thickness of the bi-layer sample. This appears to be enough to make the $T(t)$ curve non-exponential. Further increase in thickness leads to stronger deviations from exponent and more rapid reduction of the diffusion coefficient $D(t)$ (Figure \ref{different_thickness}b). However, only with $X_L = 12 $ we obtain a qualitative change at longer times and reach the $t^{-1}$ fit.

We note that the WG field pattern that we observe in all samples with cylindric symmetry at longer times is, in fact, a complex 3D oscillation  both along the round side boundary and in the vertical direction (along $Z$) between the upper and bottom sides of the cylinder.  This is indeed a whispering-gallery process caused by cylindric symmetry but not entirely a classic WG-mode well known for, e.g., solid micro-discs. There is also a constant flow of energy in the central part of the sample volume that interacts with the WG waves which is not typical for solid structures with circular symmetries. The random flow in the central part and in the bulk is a result of constant evanescent field coupling to a random network of dielectric interfaces fed by the WG-mode. Such a near-field propagation process  deserves a separate study as, to our knowledge, phenomena of this kind have not been described in literature yet.

\begin{figure}[h]%
  \centering
  \includegraphics[width=8cm]{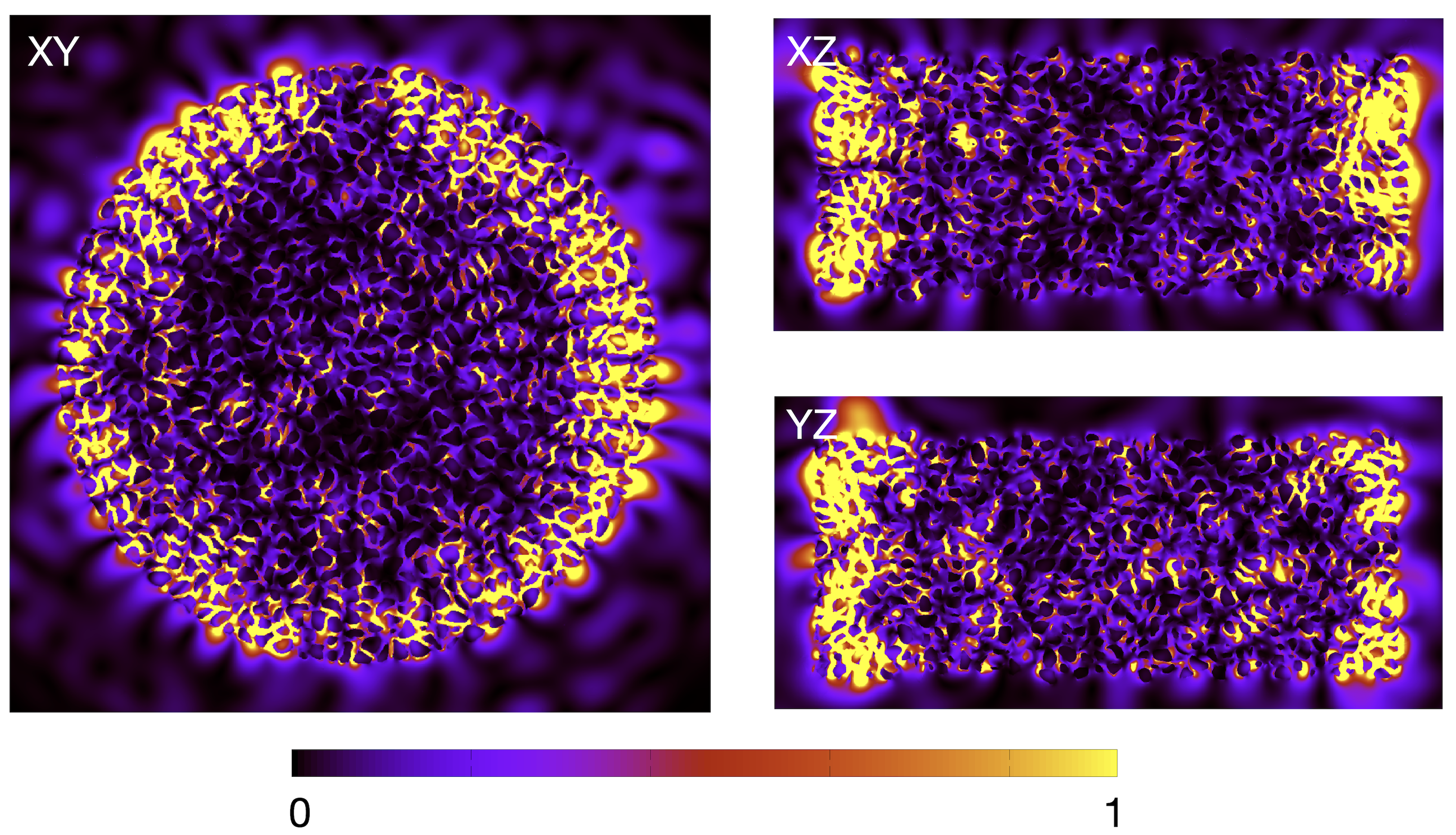}
  \caption{Cross-sections of the computational domain showing normalized near field $|E|^2$ in the sample with cylindric symmetry (Fig. \ref{fig4}a) and  $\rho=0.44$ at time  $t=25t_a$. The boundary region of the sample is intentionally oversaturated to reveal the field structure in the bulk.}
  \label{Snapshots_cylindric}
  \end{figure}

The initially randomized field component, trapped at the air–dielectric interface, exhibits a complex three-dimensional spatiotemporal energy flow. It evolves in time and the hotspot distribution eventually becomes inhomogeneous. This  becomes  apparent   in the near-field inside the $\rho=0.44$  sample, for which we observe a non-exponential tail in transmission at longer times. Figure \ref{Snapshots_cylindric} shows XY-, XZ- and YZ-cross-sections of the computational domain with the near-field intensity   for this sample at $t=25t_a$. The WG pattern along the boundary is intentionally oversaturated so that one can notice clustering of the intensity hotspots, spanning a few particle sizes, and extended dark regions in the bulk. 

  %%%%%%%%%%%%%%%%%%%%%%%%%%%%%%%%%%%%%%%%%%%%%%%%%%%%%
\subsection{ Transmission by layers with periodic boundary condition }\label{PBC_results}

In order to demonstrate that the non-exponential long-time transport persists in the absence of boundary-supported modes we did a similar series of simulations for square layers with periodic boundary conditions. However, they are not intended to reproduce the detailed crossover behavior observed in cylindric samples. 

 All model parameters, such as gaussian pulse, particle size and refractive index ($n=3.0$) remain the same, except reduced dimensions of a unit cell ($X_d =18$ along X and Y axes and $X_L = 9.6 $ along Z). We vary  volume fraction $\rho$ from 0.18 to 0.44.  PBC is another way to hinder energy decay from a finite structure. However, one should be careful and avoid phenomena like standing waves and guiding modes that may appear over long time in an artificially infinite thin layer and influence the $T(t)$ curve. 
      
      \begin{figure}[H]
  \centering
  \includegraphics[width=10cm]{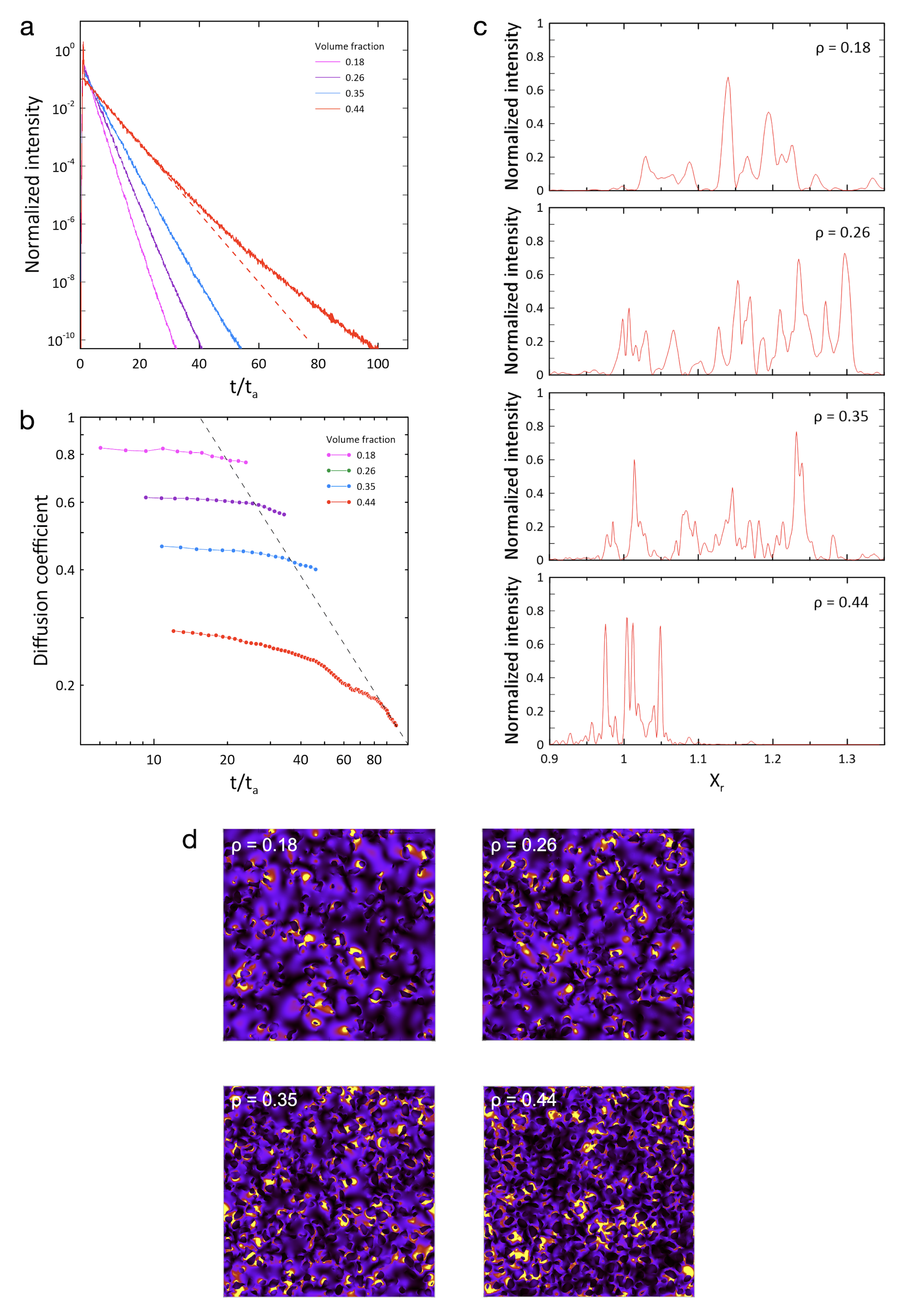}
  \caption{ a. Normalized transmission of a short pulse through periodic layers with different volume fractions (inset: $ \rho=  0.44$)  as a function of time $T(t)$. b. Diffusion coefficient $D(t)$ deduced from the $T(t)$ data. Dashed line shows a $t^{-1}$ dependence. c. Transmission spectra for different volume fractions measured at longer times and at the normalized transmission level of $T \approx 10^{-9}$. d. XY-snapshots of the near-field intensity $|E|^2$ inside layers with different volume fractions at times corresponding to the normalized transmission $T \approx 10^{-8}$.}
  \label{TtDtSpectra_periodic}
  \end{figure}
   
\begin{figure}%
  \centering
  \includegraphics[width=7cm]{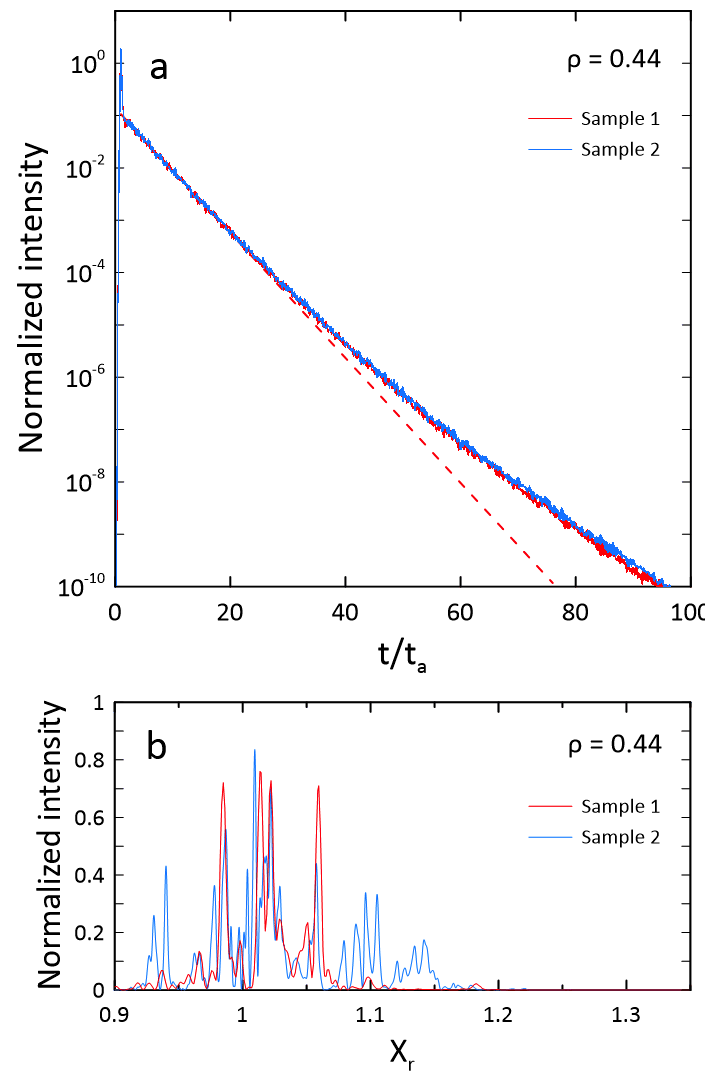}
  \caption{ Normalized transmission $T(t)$ (a) and transmission spectra (b) for two realizations of disordered layers with volume fraction $ \rho=  0.44$. }
  \label{Samples1_2}
  \end{figure}

     First of all, we need to show that such a system reveals a non-exponential tail in $T(t)$  and a time-dependent diffusion coefficient $D(t)$ at longer times. Despite the small dimensions of the unit cell and small number of particles ($N=5000$ for $\rho=0.44$)  we reach long enough observation times and see a qualitative transition in light transmission characteristics,  similar to that observed for larger finite cylindric samples. Figure \ref{TtDtSpectra_periodic} shows an emergence of a non-exponential tail in $T(t)$ at larger volume fraction ($\rho=0.44$) (a) and a time-dependent diffusion coefficient $D(t)$ (b).  $D(t)$ approaches the $t^{-1}$ regime at longer times. Transmission spectra for different $\rho$   show a transition from a number of overlapping peaks to isolated very sharp ones  (Figure \ref{TtDtSpectra_periodic}c). This indicates domination of long-lived modes whose population is responsible for the onset of $D(t) \sim t^{-1}$. The horizontal axis in the spectral plots is scaled in the units of the particle size parameter, so that the pulse central wavelength corresponds to $X_r \approx 1.15$. Such a property of the dense system is sample-independent. We additionally simulated transmission for another realization of disorder for $\rho=0.44$ and it demonstrates qualitatively similar $T(t)$ dependence and transmission spectrum (Fig \ref{Samples1_2}). The $T(t)$ curves coincide at early  times, when diffusion provides self-averaging of the field. However, they slightly diverge at long times, when a limited number of quasi-localized, realization-dependent, modes dominates. This discreteness can, in principle, lead to realization-dependent features, such as bumps or transitions in $D(t)$ as well.  Both spectra consist of sharp peaks with comparable widths concentrated in the same spectral window. Thus, we show that while WG modes can influence the detailed structure of the transmission decay in finite systems, the persistence of non-exponential behavior and other signatures under PBC indicates it is not solely attributable to such modes. 
     
     \begin{figure}%
  \centering
  \includegraphics[width=16cm]{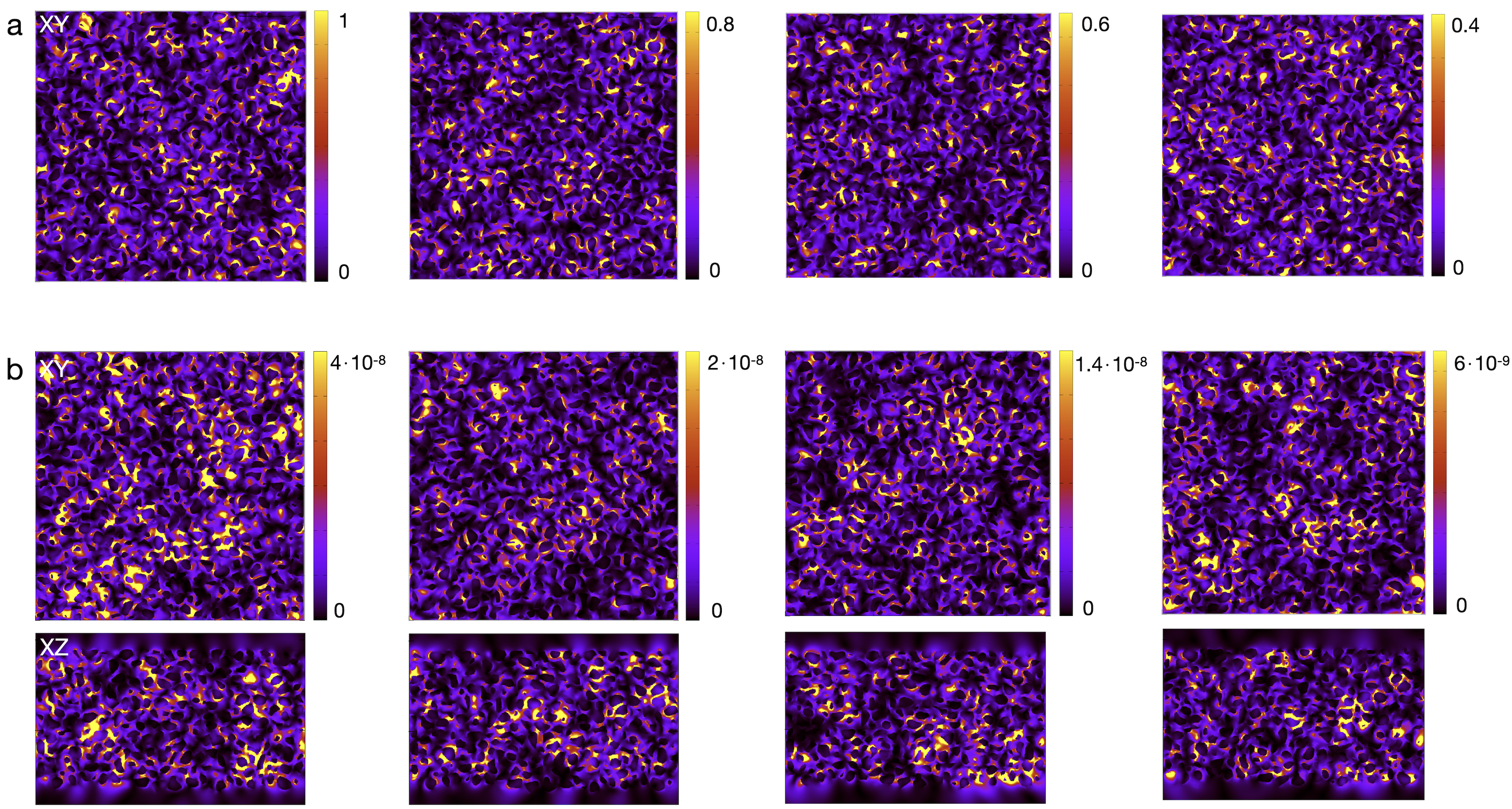}
  \caption{ Examples of the near-field intensity distributions  $|E|^2$  at different time steps,  shortly after $t = 4t_a$ (a) and at  $t > 75 t_a$ (b), when the transmission dependence $T(t)$ in Figure \ref{TtDtSpectra_periodic}a becomes significantly non-exponential in the case of volume fraction $\rho = 0.44$. }
  \label{Field_maps_periodic}
  \end{figure}

Figure \ref{TtDtSpectra_periodic}d shows   distributions of the near-field intensity in XY-plane for different volume fractions recorded at different times, when normalized transmission reaches $T(t) \approx 10^{-8}$ in each case. Here one can also notice a qualitative change. We see dipolar excitations of single particles and multiple wave scattering in a sparse structure, whereas, in a dense packing, the field energy is concentrated in small voids. Moreover, one can see clusters of hotspots and dark regions for the $\rho=0.44$ sample similarly to Fig. \ref{Snapshots_cylindric}.

As a consistency check, the transport parameter $kl^*$ can be calculated from the slope of ln$T(t)$ in its diffusive part at early times. The slope depends on the layer thickness $L$ and the diffusion coefficient $D=v_E l^* / 3$ \cite{LAGENDIJK1996}. If $s$ is the slope, fitting the diffusive part of ln$T(t)$, the diffusion coefficient can be approximately represented as $D \approx L^2 s / \pi^2$. $v_E$ is the effective wave propagation speed determined as $v_E \approx c/n_{eff}$. $n_{eff}$ can be estimated from the field maps with continuous wave propagation. It is $n_{eff}\approx 1.92$ for the volume fraction $\rho = 0.44$.  Thus, with known $D$, the transport parameter becomes $kl^* \approx 3 k n_{eff} D / c \approx 0.72$.   Similarly, for smaller volume fractions it is $kl^* \approx $ 1.32, 1.82, 2.64 for $\rho=$0.38, 0.26, 0.18, correspondingly.  We note, such an extraction of $l^*$ from the early-time ln$T(t)$ averages over the spectral content of the broadband pulse and $kl^*$ should be regarded as an effective broadband parameter with the wavenumber $k$ corresponding to the pulse  central wavelength. Monotonic increase of $kl^*$ indicates a gradual crossover from a strongly scattering regime near the localization threshold to a progressively  diffusive regime as particle density decreases. 
We note also, that experimental evidence of localization achieved at $kl^*$ exceeding 1 \cite{Mondal2019} suggests a qualitative application for the $kl^*\sim 1$ condition.    

 Now we can focus on the dense sample and track the changes in the near-field dynamics in time. Figure \ref{Field_maps_periodic} represents field snapshots taken for the $\rho=0.44$ case at different time-steps before and after the $T(t)$ curve becomes non-exponential. The upper row (Figure \ref{Field_maps_periodic}a) corresponds to the times shortly after  $t= 4 t_a$ and shows randomly scattered hotspots, indicating diffusion. XY and XZ field maps in Figure \ref{Field_maps_periodic}b are taken at  $t >  75 t_a$ when $T(t)$ is non-exponential and $D(t)$ is significantly time-dependent. At such long times field energy distribution turns into a structured pattern of interacting clusters of hotspots. These clusters can be interpreted as resonances induced by coherent propagation along highly disordered pathways. Video \ref{mode_dynamics_xz} (Supplementary material \ref{Supplement_video_PBC}) shows their dynamics. Its duration spans about 70 oscillation cycles. The clusters (as well as dark regions) evolve in time but remain stable for tens of oscillation periods which is consistent with the sharp-peak spectrum. Separate visible modes may have lifetimes of $> 20$ cycles. The lifetimes of the few sharp modes in the spectrum in Figure \ref{TtDtSpectra_periodic}d ($\rho=0.44$) can be roughly estimated from the peak widths as long as $\sim 100$ periods. These results indicate suppression of wave transport and behavior consistent with the onset of localization.
          
To quantify the long time dynamics we calculated the inverse participation ratio (IPR) for such field maps for different volume fractions and different times using the XY-cross-section data. For a 2D scalar field $z(x,y)$ it is typically defined as
  
  \begin{equation}
\textrm{IPR} =  \frac{\sum_i {z_i}^2}{ {\left( \sum_i {z_i} \right)}^2 }.
\end{equation}

$z$ is the near-field intensity in our case and IPR gives a measure of localization: lower IPR means energy spread out over many points and higher IPR implies localization at few points. 

\begin{figure}%
  \centering
  \includegraphics[width=7cm]{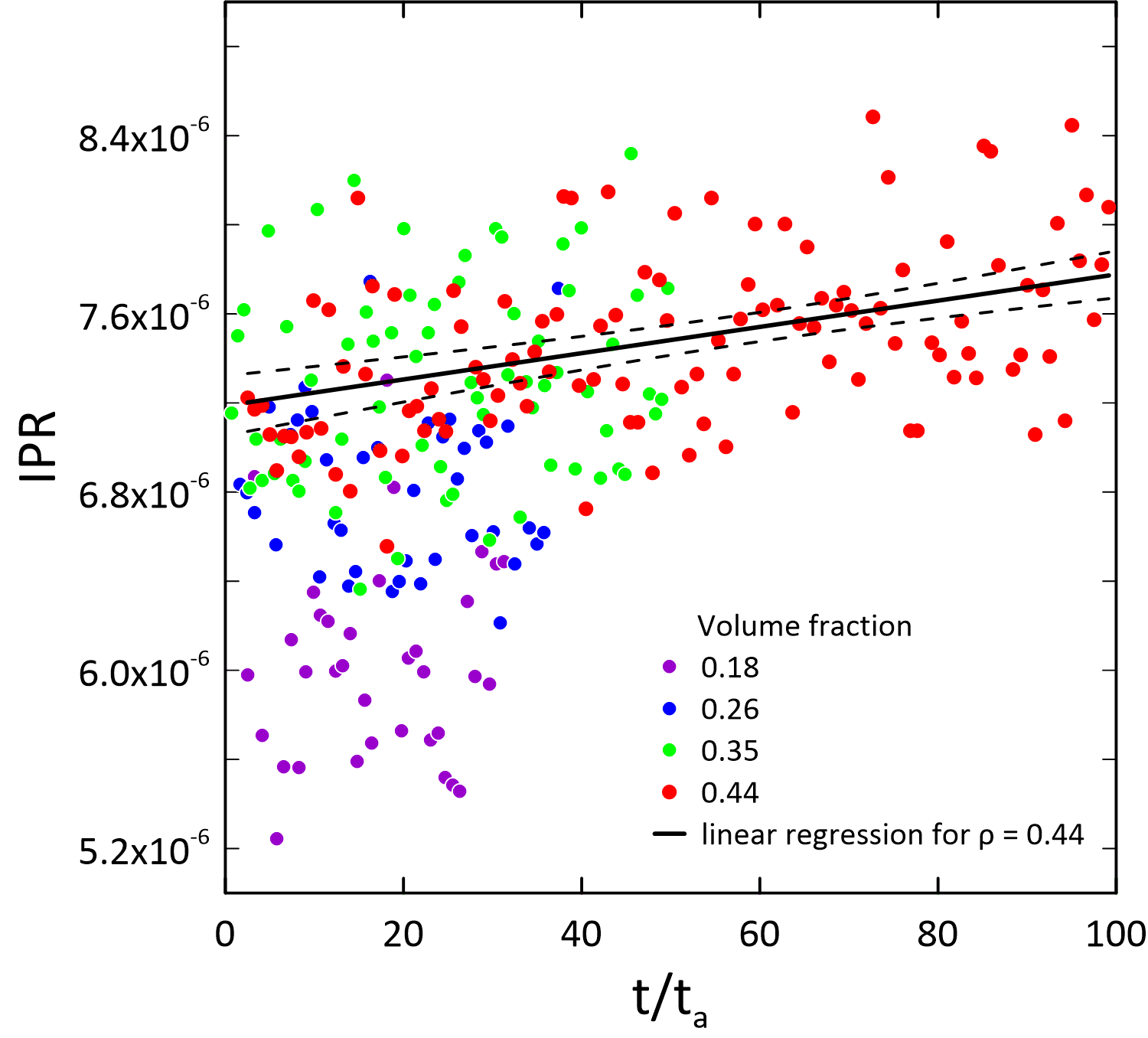}
  \caption{ Time dependence of the inverse participation ratio calculated from the intensity maps for different volume fractions. Linear regression indicates increasing energy concentration in localaized modes with time. Dashed lines show a 95\% confidence interval.}
  \label{IPR_plot}
  \end{figure}
  
The calculated IPR as a function of time   shows persistent growth for the volume fraction $\rho=0.44$ in  Figure \ref{IPR_plot},  that confirms our observation. Interestingly, this measure reflects  more uniform spread of energy in sparse samples for the same measurement period. At $\rho=0.44$,  field becomes increasingly concentrated in smaller spatial regions and the system  evolves toward stronger spatial localization over time. The contrast between bright clusters of concentrated field hotspots and dark regions persistently grows. Therefore, such a process is not stationary and the observed resonances are not eigenmodes of the system but rather a time-dependent superposition of multiple localized modes.
The excited modes can be  initially isolated but their superposition and interference between  them  creates apparent spatial connections. Thus, we can conclude that we observe a real-time evolution of the field toward AL in three dimensions.

We can give now a general interpretation of our results. Initial excitation by a short pulse creates superposition of many modes. Light explores many pathways, some of them begin to dominate due to interference while others decay. Extended modes decay faster then localized or quasi-localized ones. When the total energy of quasi-localized modes is significant enough, hotspot clusters become distinct and we obtain a slowdown in $T(t)$. After that, the system evolves toward pure Anderson modes with long-lived lifetimes, which cannot be fully reached due to the finite size of the sample and its small thickness.

The localization length can be roughly estimated  from the field images at different times as a few particle lengths or $\xi \approx 1-2 \lambda$. This is also consistent with the layer thickness test in Figure \ref{different_thickness}.  Individual modes can localize at the layer thickness $X_L = 9.6$ ($L > \xi $)  but the system is not large enough for efficient mode isolation  ($L$ is not  $\gg \xi$). Most of them  extend across a significant fraction of the layer thickness which leads to  spatial overlaps and interactions with the layer boundaries,  increasing energy leakage.  In fact, our simulation shows how Anderson localization  manifests itself in a finite system. In order to directly observe a high-quality mode isolation  we need a  layer thickness of at least $L \sim 10\xi$ or $L \sim 10-20\lambda$ with $\sim 10^5$ particles and even longer simulation times. This would require very large computing power, which is achievable, in principle, in large-scale massively parallel simulations. With our largest samples we get closer to this regime as it can be seen in Figure \ref{different_thickness}. Thus, it is not enough just to reach long observation times like we do this incorporating the WG-effect or periodic boundary conditions. Providing sufficiently large 3D volume for isolation of the disorder-induced localized modes is important.
              
The mechanism described above  can be relevant also for low-absorbing materials as it involves mostly field-interface interaction and propagation in free space. The role of the real part of the complex refractive index  becomes clear in this situation. It determines the balance between the energy transmitted by particles and groups of particles, enhancing a diffusive process, and the energy flowing through the  random network  of connected voids with long path lengths that enforces localization.

  We note, that this highly complex field propagation cannot be adequately described  by multiple light scattering in a cloud of identical point scatterers. In the last case, a  vector treatment shows that near-field coupling and   interference between the longitudinal and transverse components  suppress AL \cite{TiggelenSkipetrov2021, Skipetrov2005}. However, it has also been shown that the amplitude of the longitudinal component, specific to point scatterers, non-monotonously depends on size for spheres and conditions can even be created to minimize it \cite{Escalante2017}. In contrast to point-scatterers and spheres,  irregular particles produce inherently non-symmetric  dipolar near fields with random amplitudes determined by the particle orientation relative to the polarization of the incident wave. Consequently, the transverse and longitudinal components acquire random amplitudes, and their interference does not necessarily prevent AL. Nevertheless, the problem of the longitudinal field  remains an important question worth a separate consideration. It would be particularly interesting to study its role in  systems with different types of constituents and finely tunable spatial correlations. This can be very useful for understanding the dynamics of AL in 3D, especially since the local neighbourhood geometry is an important factor for an initial near field distribution and its subsequent evolution.

  \section{Conclusions}
  
  We have demonstrated, using large-scale full-wave simulations, that light transport in three-dimensional disordered systems of irregular dielectric particles exhibits signatures consistent with with Anderson localization under conditions of strong scattering. The transition from diffusive to localized transport is manifested by the breakdown of exponential transmission decay, the emergence of a time-dependent diffusion coefficient approaching a $t^{-1}$ scaling, and the appearance of spectrally isolated resonances with sub-unity Thouless conductance.
  
Analysis of the near-field dynamics reveals that localization develops through the formation of spatially confined, long-lived clusters of intensity hotspots. These structures arise from coherent multiple scattering and near-field coupling in densely packed media and persist over many optical cycles. The observed behavior is consistent across finite and periodic systems, indicating that it is not governed by boundary effects.

Our results highlight the importance of particle geometry, packing density, and refractive index contrast in enabling localization in dielectric systems, and show that irregular particle ensembles can overcome limitations of models based on point scatterers. The study provides a direct view of the dynamic evolution toward localization in three dimensions and establishes a computational framework for investigating strongly disordered photonic media beyond the regime of conventional multiple scattering. It also motivates experimental efforts to identify Anderson localization signatures in realistic disordered media with refractive indices not exceeding $n = 3.0$.

\begin{acknowledgments}
  The authors gratefully acknowledge the computing time granted by the Paderborn Center for Parallel Computing (PC$^2$).
\end{acknowledgments}

\bibliography{bibliography}% Produces the bibliography via BibTeX.

\pagebreak 

%Supplement
%\appendix

\renewcommand{\thesection}{\Roman{section}}
\renewcommand{\thefigure}{S\arabic{figure}}
\renewcommand{\thevideo}{S\arabic{video}}
\setcounter{figure}{0}
%\counterwithin{figure}{section}

%%%%%%%%%%%%%%%%%%%%%%%%%%%%&%%%%%%%%%%%%%%%%%%%%%%%%%
\section*{Supplementary material}

\section{CST check  for  a sample with reduced dimensions}\label{CST_check}

\begin{figure}
  \centering
  \includegraphics[width=5cm]{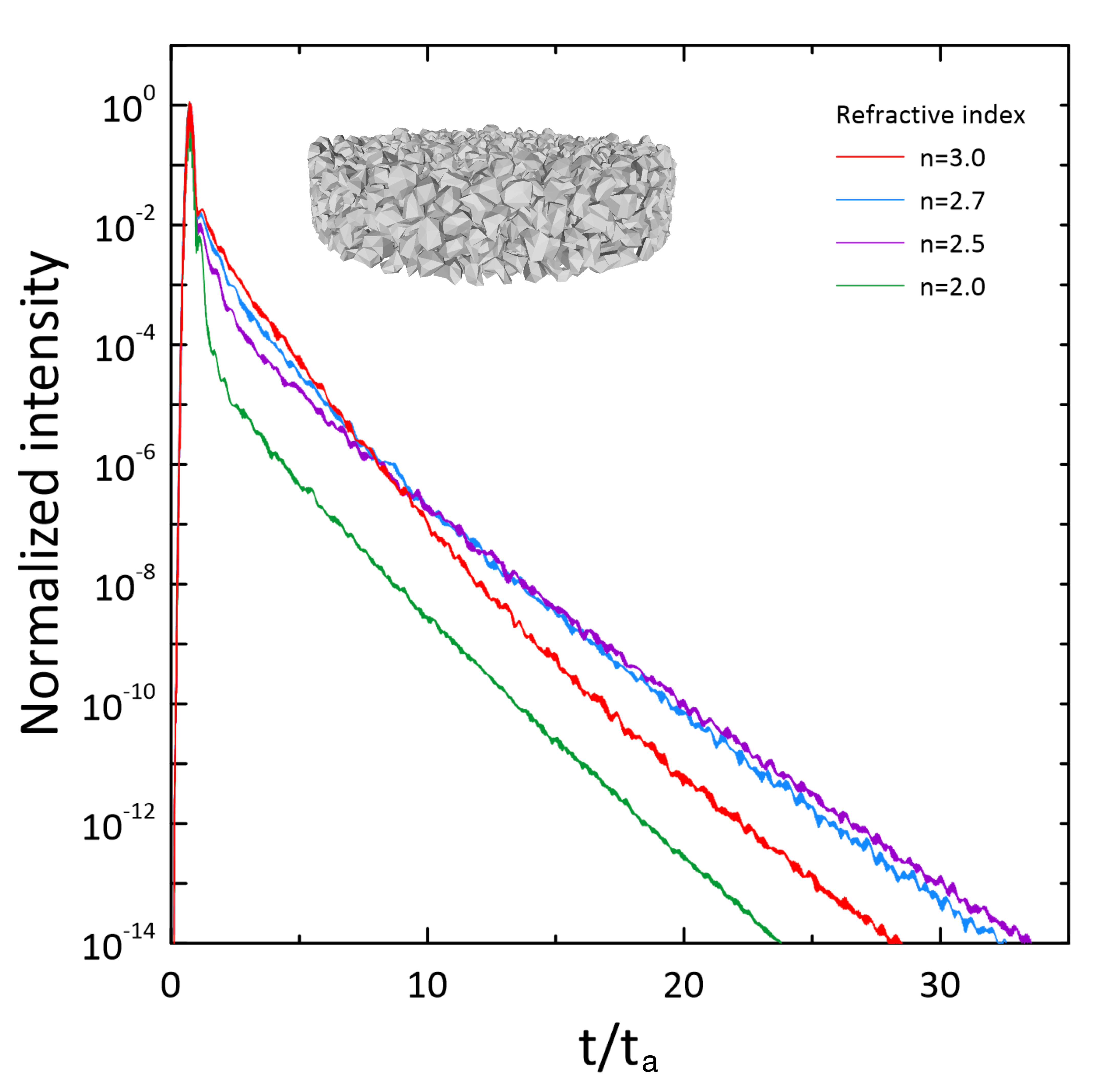}
  \caption{ Normalized transmission of a short pulse  through a small layer of 2000 particles packed with volume fraction $ \rho=  0.5$ (inset) and with dimensions  $X_R= 15$ and $X_L = 3$ as a function of time $T(t)$ at different refractive indices. Simulations are done with the  \textit{CST Microwave Studio} software package.}
  \label{CST_plot}
  \end{figure}
 
Here we do test simulations using the  \textit{CST Microwave Studio} software package. It is based on the Time Domain Finite Integration Technique, so, one can qualitatively benchmark our DGTD results with a different numerical method and a different electromagnetic solver. 

   For the computations with the  CST FIT solver we take a cylindric sample with thickness $X_L = 3$ (a four-particle thick layer) and reduced diameter  $X_R = 15$. Due to parallelization restrictions of CST MWS we need to simplify the target system.  With the particle size $X_r=1$  and  the volume fraction  $\rho = 0.5$ the number of particles becomes relatively small in this case, N=2000. With fixed sample parameters we vary  the material refractive index $n$ from 2.0 to 3.0. For the CST computations we use a polyhedral description of the model sample and, then, a hexahedral mesh is generated by the software internally. The electromagnetic problem is solved in time domain simulating propagation of the short plane-wave pulse, similarly to our DGTD method setup.
  
   The CST results are shown in in Figure \ref{CST_plot} and are qualitatively similar to the $T(t)$ refractive index dependences in Figure \ref{fig5}a obtained for a  larger system with the DGTD method.

  %%%%%%%%%%%%%%%%%%%%%%%%%%%%%%%%%%%%%%%%%%%%%%%%%%%
\section{Early-time near-field dynamics  }\label{Supplement_video}

In this section we present animated visualization of a short pulse propagation in a thick layer with a large number of particles, shown in Figure \ref{fig1}a. Video \ref{short_pulse_video} demonstrates propagation of a short focused pulse for the same model parameters ($X_r = 1.1$, $n=3$). It represents an XZ-cross-section of the computational domain with the near-field intensity distribution. One can see a a CA-like behavior of the  field hotspots in the sub-wavelength voids between the closely neighboring particles. The wave packet is partially preserved along the propagation path. Its effective wavelength becomes smaller,  $\lambda_{eff} \approx 0.6\lambda_{inc} $, increasing the effective size of particles to $X_{r,eff}  \approx 1.8 $. Nevertheless, the condition $ kr \sim 1$ is preserved for the most of particles and inter-particle voids.

%Video 1
\begin{video}
  \centering
  \href{https://arxiv.org/src/2312.14393v1/anc/A1transverse_localized_short_xz.mp4}{\includegraphics[width=4cm]{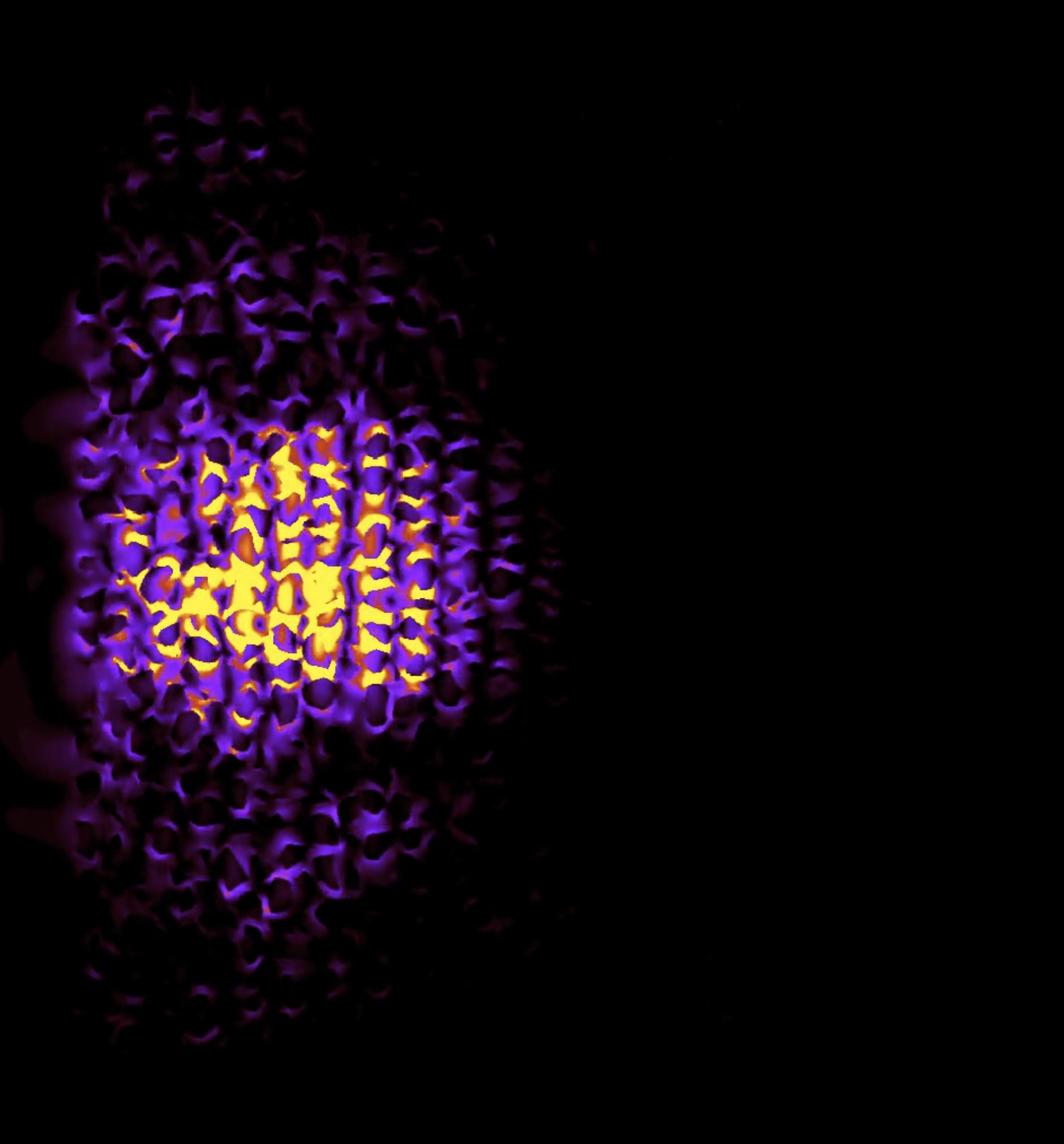}}
  \setfloatlink{https://arxiv.org/src/2312.14393v1/anc/A1transverse_localized_short_xz.mp4}%
  \caption{\label{short_pulse_video}
  Propagation of a short focused pulse in a dense layer shown in Figure \ref{fig1}a. %(A1\_transverse\_localized\_short\_xz.mp4)
 }%
\end{video}

Video \ref{short_plane_wave_video} represents the dynamics of the randomized near field   excited in the same sample by a plane-wave short pulse at very early times after $2t_a$.  One can notice that the decay of hotspots is not monotonic in each location. They demonstrate different behavior in such a disordered system already at earlier times: those that are excited directly by the transmitted pulse decay while others arise in new locations. The decay rate is also different for different hotspots, i.e. they may have different life times. Some are decaying within a few oscillations while others form clusters and are much more stable, surviving up to $\approx 10$ cycles. Non-monotonic decay and enhancement with time may also take place in the same location, which proves a high geometrical complexity of the paths and mode population. If the energy of a path is leaked at the sample boundary, it is lost in the total decay. A leakage in the interior transfers energy to new paths with similar spatial scales and degrees of complexity, forming a complex percolation graph.  Evanescent coupling plays a crucial role here. Small dielectric particles act as resonant scatterers and the evanescent field tails from their surfaces overlap with neighboring particles. Such overlap is phase-sensitive and provides coherent pathways in a dense system.
     
      %Video 2
\begin{video}
  \centering
  \href{https://arxiv.org/src/2312.14393v1/anc/A2dynamics_after_transmission_xz.mp4}{\includegraphics[width=4cm]{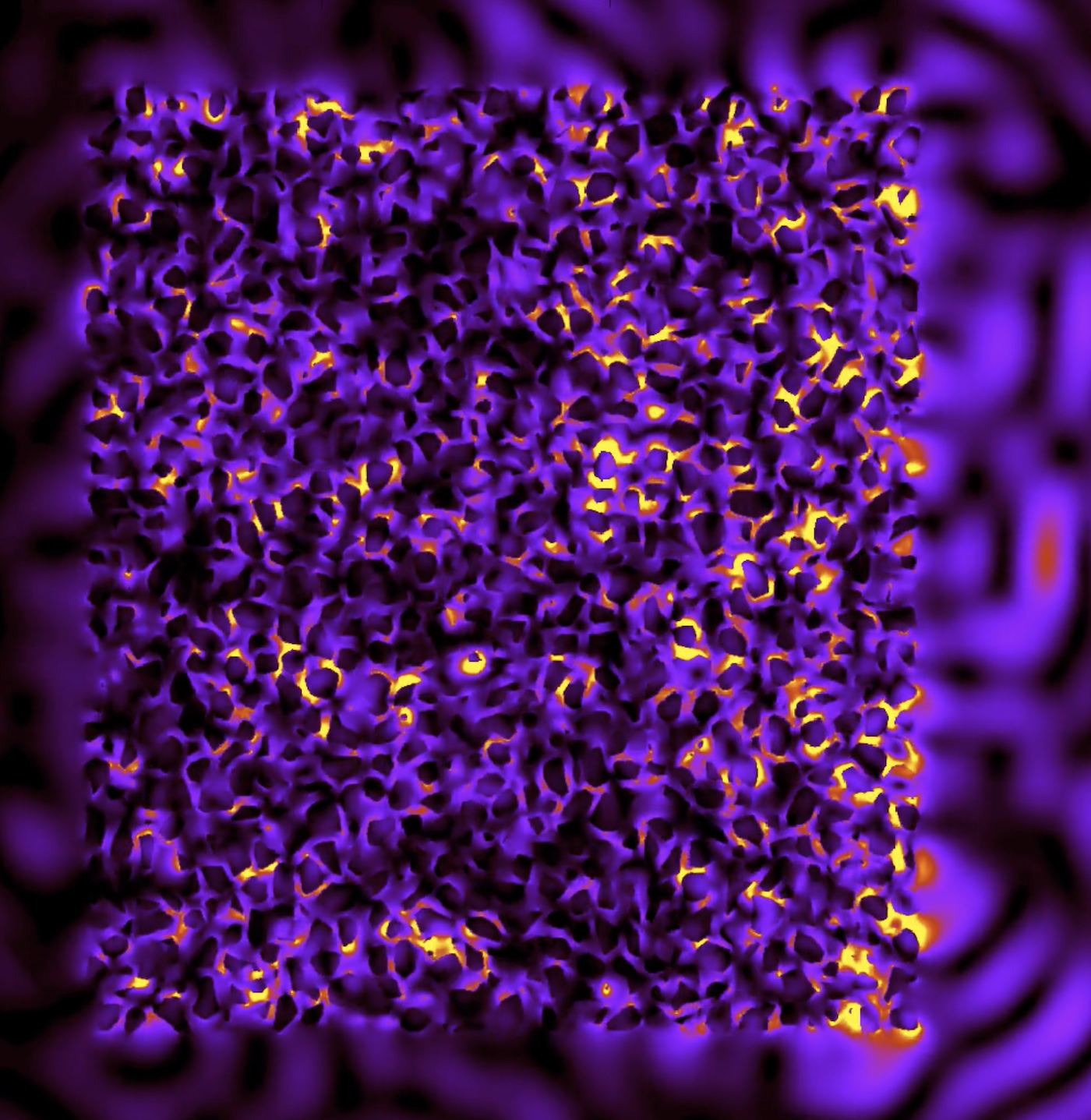}}
  \setfloatlink{https://arxiv.org/src/2312.14393v1/anc/A2dynamics_after_transmission_xz.mp4}%
  \caption{\label{short_plane_wave_video}
   Dynamics of the near field for a  period of time after $2t_a$ excited by a short plane wave pulse in a dense layer with $\rho=0.48$ and $n=3.0$ shown in Figure \ref{fig1}a.% (A3\_dynamics\_after\_transmission\_xz.mp4)
 }%
\end{video}

\section{Late-time near-field dynamics for a periodic sample}\label{Supplement_video_PBC}
   
Video \ref{short_plane_wave_video} demonstrates persistent oscillations of the quasi-localized modes in a system with peridoc boundaries in transverse directions at late times. Non-propagating pinned clusters of hotspots indicate suppression of wave transport.

 %Video 3
\begin{video}
  \centering
  \href{https://drive.google.com/file/d/1JC4m16wSRwCg5Bl-BwbjG-NyeSPUbHjM/view?usp=sharing}{\includegraphics[width=8cm]{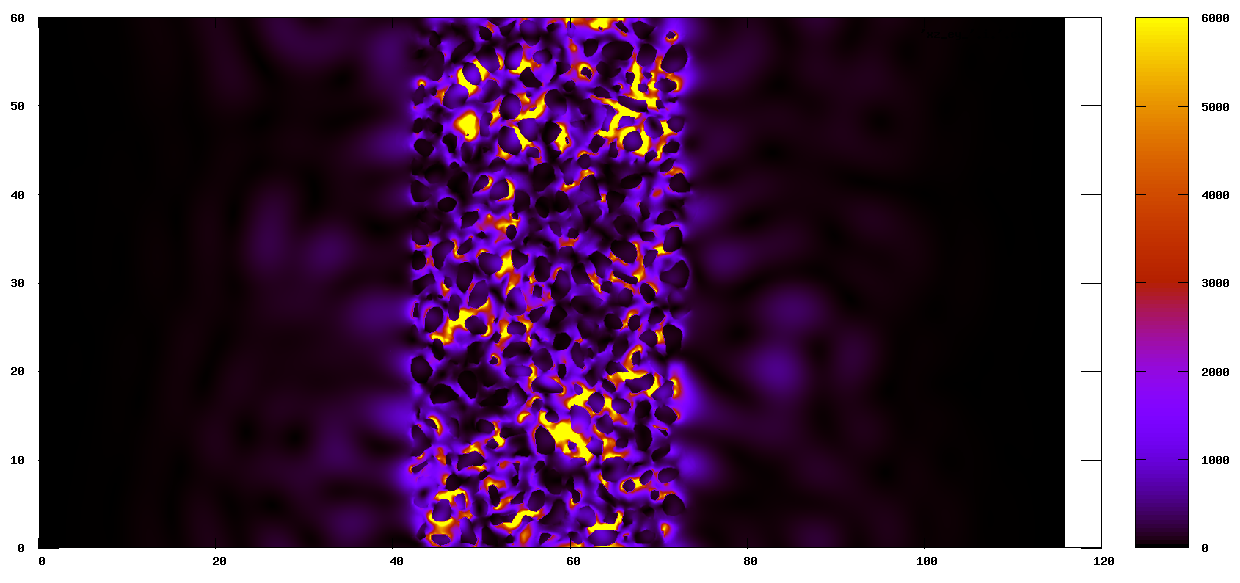}}
  \setfloatlink{https://arxiv.org/src/2312.14393v1/anc/A3modes_xz.mp4}%
  \caption{\label{mode_dynamics_xz}
   Dynamics of the near field intensity $|E|^2$ in  XZ-plane for a  period of time after $t =  75 t_a$ showing stable disorder-induced modes in a dense layer with $\rho=0.44$ and periodic boundary conditions.% (A3\_dynamics\_after\_transmission\_xz.mp4)
 }%
\end{video}

\end{document}